\documentclass[a4paper,12pt]{article}

\usepackage{latexsym}
\usepackage{graphics}
\usepackage{amsmath}

\begin{document}

\baselineskip = 0.33 in
\topmargin= -15mm
\textheight= 230mm

\begin{center}
{\Large  {\bf Ground state property of Bose-Einstein gas for arbitrary 
power low interaction}}
\end{center}

\vspace{0.5cm}

\begin{center}
M. Hiramoto\footnote{e-mail: hiramoto@phys.ge.cst.nihon-u.ac.jp}\\
College of Science and Technology, Nihon University,\\
 Funabashi, Chiba 274-8501, Japan
\end{center}

\vspace{0.5cm}

\begin{center}
\section*{\large Abstract} 
\end{center}

We study Bose-Einstein gas for an arbitrary power low interaction 
$C_{\alpha}r^{-\alpha}$. This is done by the Hartree Fock Bogoliubov (HFB) 
approach at $T \le T_{c}$ and the mean field approach at $T>T_{c}$. Especially, 
we investigate the ground state property of Bose gas interacting through 
the Van der Waals $-C_{6}r^{-6}$ plus $C_{3}r^{-3}$ interactions. We show that 
the ground state under this interaction is stable if the ratio of coupling 
constants is larger than that of the critical curve. We find that the 
$C_{3}r^{-3}$ term plays an important role for the stability of the ground 
state when the density of atoms becomes sufficiently large at low temperature. 
Further, using the numerical values of $C_{3}$ and $C_{6}$, we confirm that 
the ground state of alkali atoms are stable.

\vspace{0.5cm}
PACS numbers: 03.75.Fi, 05.30.Jp, 32.80.Pj, 51.30.+i \par

\newpage

\section{Introduction}
Bose-Einstein condensation (BEC) was first observed in dilute ultracold alkali 
atoms of rubidium \cite{and}, lithium \cite{bra} and sodium \cite{dav}. 
Furthermore, Mewes et al. and Ensher et al. measured the condensate fraction 
and the energy of rubidium atoms  \cite{mew,ens}. It is shown that the 
transition temperature $T_{0}$ is shifted not only by the interaction effect 
but also by the finite size effect by a few percent, where $T_{0}$ denotes the 
transition temperature of the noninteracting Bose gas within the external field 
in the thermodynamic limit. This means that both the interaction and the finite 
size effects play an important role in the real Bose gases. In fact, the number 
of trapped atoms is typically $N \le 10^7$, and this may not be 
sufficiently large to take the thermodynamic limit. 
\\
On the other hand, these experimental findings have stimulated much interest 
in the theory of the interacting Bose gas. Since BEC occurs when atoms are 
dilute and cold, we can treat the interaction of atoms as the two-body 
interaction. In this case, the s-wave scattering length characterizes the 
strength of the two-body interaction. Under these conditions, the Gross 
Pitaevskii (GP) equation can well describe the behavior of the interacting Bose 
gas at zero temperature \cite{gro,pit}. This is a mean field approach for the 
order parameter associated with the condensate. Using the GP equation, several 
authors studied the ground state and the excitation properties of the 
condensate \cite{dgp}.
\\
To study BEC at finite temperature, the GP equation at zero temperature was 
extended by Griffin \cite{gri}. It is called the Hartree Fock Bogoliubov (HFB) 
theory. In particular, Popov approximation to the HFB theory has been employed 
to explain the experimental results \cite{gioa}.
\\
In the realistic point of view, BEC experiments are carried out in magnetic 
traps. In this situation, spin-polarized atoms interact through a triplet 
potential. As long as atoms remain polarized, they cannot form molecules. For 
alkali atoms, the triplet potential has many bound states which allow them to 
recombine into molecules. Since this recombination can only occur in a 
three-body scattering, it cannot occur for sufficiently low density of atoms. 
Thus, the two-body scattering is dominant. Therefore, spin-polarized atoms can 
remain the gas through dipole two-body scattering which flip the spin to 
untrapped state, and then atoms can produce BEC. In this sense, it is of 
particular importance to find a way to investigate BEC with more realistic 
interactions. 
\\
In this paper, we study BEC for an arbitrary power low interaction 
$C_{\alpha}r^{-\alpha}$ which is more realistic than the ordinary contact 
interaction ($\delta$-function). This interaction plays an important role for 
alkali atoms \cite{dal}. The atomic interaction $V$ of alkali atoms can be 
approximately written as \cite{mwjk}
\begin{eqnarray}
V &=& V_{c}+V_{d}+V_{hf}+V_{Z}+V_{so},
\end{eqnarray}
where $V_{c}$ is the central force of the interaction. Further, $V_{c}$ can be 
written in terms of the electron exchange interaction $V_{ex}$ and the 
dispersion force $-C_{6}r^{-6}-C_{8}r^{-8}-C_{10}r^{-10}\cdots$. The 
magnetic-dipole interaction $V_{d}$ and the hyperfine interaction $V_{hf}$ 
behave like $C_{3}r^{-3}$, where the coupling constant $C_{3}$ is the spin 
part of the interaction. The last two terms, $V_{Z}$ and $V_{so}$ represent 
the Zeeman interaction and the spin-orbit interaction, respectively. These 
interactions are important at high density.
\\
Here, we study Bose gas interacting through the Van der Waals 
$-C_{6}r^{-6}$ plus $C_{3}r^{-3}$ interactions. We show that the ground state 
under this interaction is stable if the ratio of the coupling constants 
is larger than that of the critical curve. We find that the $C_{3}r^{-3}$ term 
plays an important role to stabilize the ground state when the density of 
atoms becomes large at low temperature. Using the numerical values of 
$C_{3}$ and $C_{6}$, we confirm that the ground state of alkali atoms are 
stable.
\\
This paper is organized in the following way. In the next section, we derive 
the effective Hamiltonian for the interacting Bose gas. Then, in section 3, we 
study the ground state stability for the Van der Waals plus $C_{3}r^{-3}$ 
interactions. Section 4 summarizes what we have clarified in this 
paper.

\section{The Effective Hamiltonian for Interacting Bose Gas}
In this section, we derive the effective Hamiltonian for the interacting Bose 
gas. Since Bose gas dramatically changes its behavior below the transition 
temperature $T_c$, we derive the effective Hamiltonian with the HFB approach 
at $T\le T_c$. On the other hand, we derive the effective Hamiltonian with the 
mean field approach at $T>T_c$.
\\
The Hamiltonian for the interacting Bose gas confined in a harmonic oscillator 
potential can be written as
\begin{eqnarray}
\hat{H} & = & \int d{\bf r}\hat{\Psi}^{\dagger}({\bf r})
\left[-\frac{\hbar^2}{2m}\nabla^2+\frac{1}{2}m\omega r^{2}\right]
\hat{\Psi}({\bf r})
\nonumber \\
& & +\frac{1}{2}\int d{\bf r}_{1}d{\bf r}_{2}
\hat{\Psi}^{\dagger}({\bf r}_{1})\hat{\Psi}^{\dagger}({\bf r}_{2})
V(|{\bf r}_{1}-{\bf r}_{2}|)
\hat{\Psi}({\bf r}_{2})\hat{\Psi}({\bf r}_{1}),
\end{eqnarray}
where $\hat{\Psi}({\bf r})$ is the boson field operator. The two-body atomic 
interaction $V(|{\bf r}_{1}-{\bf r}_{2}|)$ is given as
\begin{eqnarray}
V(|{\bf r}_{1}-{\bf r}_{2}|) & = & \frac{C_{\alpha}}{|{\bf r}_{1}-
{\bf r}_{2}|^{\alpha}}.
\end{eqnarray}
Now, we write down the second quantized Hamiltonian for Eq. (2) \cite{row}. 
The boson field operator for the ideal system can be expanded by plane waves, 
but for the general case, the corresponding field operator is a sum over all 
normal modes \cite{fet}
\begin{eqnarray}
\hat{\Psi}({\bf r}) & = & \sum_{{\bf \nu}}\hat{a}_{{\bf \nu}}\chi_{{\bf \nu}}
({\bf r}),
\end{eqnarray}
when $\chi_{\bf \nu}(\bf r)$'s are any complete set of normalized 
single-particle wave function, and $a_{{\bf \nu}}$ is a bosonic annihilation 
operator for the single-particle state ${\bf \nu}$. From Eq. (4), the 
Hamiltonian can be written as
\begin{eqnarray}
\hat{H} & = & 
\sum_{{\bf \nu \nu'}} T_{{\bf \nu \nu'}}\hat{a}^{\dagger}_{{\bf \nu}}
\hat{a}_{{\bf \nu'}}+\frac{1}{2}\sum_{{\bf \nu \lambda \nu' \lambda'}}
V_{\bf{\nu' \lambda'}}^{\bf{\nu \lambda}}
\hat{a}^{\dagger}_{{\bf \nu}}\hat{a}^{\dagger}_{{\bf \lambda}}
\hat{a}_{{\bf \nu'}}\hat{a}_{{\bf \lambda'}},
\end{eqnarray}
with
\begin{eqnarray}
T_{{\bf \nu \nu'}} & = & \int d{\bf r} \chi^{\dagger}_{{\bf \nu}}({\bf r})
\left[-\frac{\hbar^2}{2m}\nabla^2+V_{\rm{ext}}({\bf r})\right]
\chi_{{\bf \nu'}}({\bf r}), \\
V_{\bf{\nu' \lambda'}}^{\bf{\nu \lambda}} 
& = & \int d{\bf r}_{1}d{\bf r}_{2}\chi^{\dagger}_{{\bf \nu}}({\bf r}_{1})
\chi^{\dagger}_{{\bf \lambda}}({\bf r}_{2})V(|{\bf r}_{1}-{\bf r}_{2}|)
\chi_{{\bf \nu'}}({\bf r}_{2})\chi_{{\bf \lambda'}}({\bf r}_{1}).
\end{eqnarray}
Here, we expand $V(|{\bf r}_{1}-{\bf r}_{2}|)$ in terms of the Legendre 
polynomial 
$P_{l}$\begin{eqnarray}
V(|{\bf r}_{1}-{\bf r}_{2}|) &=& 
\sum_{l=0}^{\infty}v_{l}(r_{1},r_{2})P_{l}(\cos\theta_{12}),\nonumber \\
&=& \sum_{l=0}^{\infty}\sum_{m=-l}^{l}\frac{4\pi}{2l+1}v_{l}(r_{1},r_{2})
Y_{l,m}^{\ast}(\theta_{1},\varphi_{1})Y_{l,m}(\theta_{2},\varphi_{2}),
\end{eqnarray}
where $v_{l}(r_{1},r_{2})$ is given by
\begin{eqnarray}
v_{l}(r_{1},r_{2}) &=& \frac{2l+1}{2}\int_{-1}^{1}dt\ 
V(|{\bf r}_{1}-{\bf r}_{2}|)P_{l}(t).
\end{eqnarray}
In the case of the power low potential $ V(|{\bf r}_{1}-{\bf r}_{2}|)=
C_{\alpha}|{\bf r}_{1}-{\bf r}_{2}|^{-\alpha}$, we can integrate $v_{l}(r_{1},
r_{2})$;
\begin{eqnarray}
v_{l}(r_{1},r_{2}) &=& C_{\alpha}\frac{\Gamma(\frac{1}{2})\Gamma
(l+\frac{\alpha}{2})}{\Gamma(l+\frac{1}{2})\Gamma(\frac{\alpha}{2})}
\frac{r_{2}^{l}}{r_{1}^{\alpha+l}}\ 
_{2}F_{1}\left(\frac{\alpha}{2}+l,\frac{\alpha}{2}-\frac{1}{2}
,l+\frac{3}{2},\frac{r_{2}^{2}}{r_{1}^{2}}\right),
\end{eqnarray}
for $r_{1}>r_{2}$, and
\begin{eqnarray}
v_{l}(r_{1},r_{2}) &=& C_{\alpha}\frac{\Gamma(\frac{1}{2})\Gamma
(l+\frac{\alpha}{2})}{\Gamma(l+\frac{1}{2})\Gamma(\frac{\alpha}{2})}
\frac{r_{1}^{l}}{r_{2}^{\alpha+l}}\ 
_{2}F_{1}\left(\frac{\alpha}{2}+l,\frac{\alpha}{2}-\frac{1}{2}
,l+\frac{3}{2},\frac{r_{1}^{2}}{r_{2}^{2}}\right),
\end{eqnarray}
for $r_{2}>r_{1}$, where $_{2}F_{1}(a,b,c,x)$ denotes hypergeometric function. 
Thus, we obtain the second quantized Hamiltonian for Bose gas interacting 
through the power low interaction. From now on, we present the mean field 
Hamiltonian and the HFB Hamiltonian.
\\
\subsection{Mean Field Theory}
First, we present the mean field theory for the interacting Bose gas 
\cite{lan}. This theory is particularly valid for $T> T_{c}$ \cite{sg}. But 
for $T\le T_{c}$, this theory is inadequate, since the low energy part of the 
excited states plays an important role at low temperature. This effective 
Hamiltonian $\hat{H}_{eff}$ is given by the diagonal part of the Hamiltonian 
(5), and can be written as 
\begin{eqnarray}
\hat{H}_{eff} & = & \sum_{{\bf \nu}} T_{{\bf \nu\nu}}\hat{n}_{{\bf \nu}}
+\frac{1}{2}\sum_{{\bf \nu\nu'}}v_{{\bf \nu\nu'}}
\hat{n}_{{\bf \nu}}\hat{n}_{{\bf \nu'}},
\end{eqnarray}
where $n_{{\bf \nu}}=a^{\dagger}_{{\bf \nu}}a_{{\bf \nu}}$ and 
\begin{eqnarray}
v_{{\bf \nu\nu}} &=& \int d{\bf r}_{1}d{\bf r}_{2}|\chi_{{\bf \nu}}
({\bf r}_{1})|^{2}V(|{\bf r}_{1}-{\bf r}_{2}|)|\chi_{{\bf \nu}}
({\bf r}_{2})|^{2},
\end{eqnarray}
for ${\bf \nu}={\bf \nu'}$ and
\begin{eqnarray}
v_{{\bf \nu\nu'}} &=& \int d{\bf r}_{1}d{\bf r}_{2}|\chi_{{\bf \nu}}
({\bf r}_{1})|^{2}V(|{\bf r}_{1}-{\bf r}_{2}|)|\chi_{{\bf \nu'}}
({\bf r}_{2})|^{2}
\nonumber\\&&
+\int d{\bf r}_{1}d{\bf r}_{2}\chi^{\dagger}_{{\bf \nu}}({\bf r}_{1})
\chi^{\dagger}_{{\bf\nu'}}({\bf r}_{2})V(|{\bf r}_{1}-{\bf r}_{2}|)
\chi_{{\bf \nu}}({\bf r}_{2})\chi_{{\bf \nu'}}({\bf r}_{1})
\end{eqnarray}
for ${\bf \nu}\neq{\bf \nu'}$. We rewrite Eq. (12) in the following way
\begin{eqnarray}
\hat{H}_{eff} & = & \hat{H}_{0}+\hat{H}',
\end{eqnarray}
with
\begin{eqnarray}
\hat{H}_{0} & = & \sum_{{\bf \nu}}\tilde{E}_{{\bf \nu}}\hat{n}_{{\bf \nu}}-
\frac{1}{2}\sum_{{\bf \nu\nu'}}v_{{\bf \nu\nu'}}
\rho_{\bf \nu}\rho_{\bf \nu'},\\
\hat{H}' & = & \frac{1}{2}\sum_{{\bf \nu\nu'}}v_{{\bf \nu\nu'}}
(\hat{n}_{{\bf \nu}}-\rho_{\bf \nu})
(\hat{n}_{{\bf \nu'}}-\rho_{\bf \nu'}),\\
\tilde{E}_{{\bf \nu}} & = & T_{{\bf \nu\nu}}+\sum_{{\bf \nu'}}v_{{\bf \nu\nu'}}\rho_{\bf \nu'}.
\end{eqnarray}
Here, we minimize $\hat{H}'$ by taking $\rho$'s in the following way
\begin{eqnarray}
\rho_{\bf \nu} = \langle n_{{\bf \nu}}\rangle = 
\frac{{\rm{Tr}}\ n_{{\bf \nu}}\ e^{-(\hat{H}_0-\mu \hat{N})/k_B T}}
{{\rm{Tr}}\ e^{-(\hat{H}_0-\mu \hat{N})/k_B T}} = 
\frac{1}{e^{(\tilde{E}_{{\bf \nu}}-\mu )/k_B T}-1}. 
\end{eqnarray}
Then, the thermodynamic potential $\Omega=-pV$ can be given as
\begin{eqnarray}
\Omega & = & -k_{B}T\ln {\rm Tr}\ e^{-(\hat{H}_{0}-\mu \hat{N})/k_{B}T},
 \nonumber \\
& \simeq & -\frac{1}{2}\sum_{{\bf \nu\nu'}}v_{{\bf \nu\nu'}}
\langle n_{{\bf \nu}}\rangle\langle n_{{\bf \nu'}}\rangle
+k_{B}T\sum_{{\bf \nu'}}\ln{(1-e^{-(\tilde{E}_{{\bf \nu'}}-\mu)/k_{B}T})}.
\end{eqnarray}
We note that Eq. (19) is also determined by the condition 
\begin{eqnarray}
\left(\frac{\partial \Omega}{\partial \langle n_{{\bf \nu}}\rangle}\right)
_{T,\mu,V,\langle n_{{\bf \nu'}}\rangle\ne \langle n_{{\bf \nu}}\rangle}
= -\sum_{{\bf \nu'}}v_{{\bf \nu\nu'}}\langle n_{{\bf \nu'}}\rangle
+\sum_{{\bf \nu'}}v_{{\bf \nu\nu'}}\frac{1}{e^{(\tilde{E}_{{\bf \nu'}}-\mu )/k_B T}-1}=0.
\end{eqnarray}
This condition means that the thermodynamic potential $\Omega$ is an extremum 
with respect to any change of $T$, $V$ and $\mu$ in a state of thermal 
equilibrium. 
\\
\subsection{HFB Theory}
Now, we present the HFB theory. This theory is reliable for the description 
of the low temperature behavior of the Bose gas. In this theory, the operators 
$a_{0}$ and $a_{0}^{\dagger}$ are replaced by the c-number 
$a_{0}, a_{0}^{\dagger} \approx \sqrt{N_{0}}$. Then, the HFB Hamiltonian 
$\hat{K}=\hat{H}-\mu \hat{N}$ is given by
\begin{eqnarray}
\hat{K} &=& (T_{00}-\mu)N_{0}+\frac{1}{2}V^{00}_{00}N_{0}^{2}\nonumber\\
& &
+\sum_{{\bf \nu\nu'}\neq 0}\left\{T_{\bf{\nu\nu'}}-\mu\delta_{\bf{\nu\nu'}}
+2N_{0}V^{0\nu}_{0\nu'}+2n'_{\bf{\lambda}}V^{\bf{\lambda\nu}}_{\bf{\lambda\nu'}}
\right\}a_{\bf{\nu}}^{\dagger}a_{\bf{\nu'}}\nonumber\\
& &+\frac{N_{0}}{2}\sum_{{\bf \nu\nu'}\neq 0}
V^{00}_{\bf{\nu\nu'}}(a_{\bf{\nu}}a_{\bf{\nu'}}+
a_{\bf{\nu}}^{\dagger}a_{\bf{\nu'}}^{\dagger}),
\end{eqnarray}
where $n'_{\bf{\lambda}}$ is the expectation value of the non-condensate 
density of particles, and we eliminate the lower power of $N_{0}$ and terms 
which are proportional to $a_{\bf{\nu}}$ and $a_{\bf{\nu}}^{\dagger}$. Now, we 
define new boson operators by Bogoliubov transformation
\begin{eqnarray}
c_{\bf{\nu}} &=& u_{\bf{\nu}}a_{\bf{\nu}}+v_{\bf{\nu}}a^{\dagger}_{\bf{\nu}}, \\
c^{\dagger}_{\bf{\nu}} &=& 
u_{\bf{\nu}}a^{\dagger}_{\bf{\nu}}+v_{\bf{\nu}}a_{\bf{\nu}},
\end{eqnarray}
where $u_{\bf{\nu}}$ and $v_{\bf{\nu}}$ satisfy 
$u_{\bf{\nu}}^{2}-v_{\bf{\nu}}^{2}= 1$. Then, we can write the Hamiltonian in 
terms of the new operator $c_{\bf{\nu}}$
\begin{eqnarray}
\hat{K} &=& (T_{00}-\mu)N_{0}+\frac{1}{2}V^{00}_{00}N_{0}^{2}
+\sum_{{\bf \nu\nu'}\neq 0}\left(\varepsilon_{\bf{\nu\nu'}}v^{2}_{\bf{\nu}}
-N_{0}V^{00}_{\bf{\nu\nu'}}
u_{\bf{\nu}}v_{\bf{\nu'}}\right)
\nonumber \\
& &
+\sum_{{\bf\nu\nu'}\neq 0}
\biggl[\left\{\varepsilon_{\bf{\nu\nu'}}
(u_{{\bf \nu}}u_{\bf{\nu'}}+v_{\bf{\nu}}v_{\bf{\nu'}})
-N_{0}V^{00}_{\bf{\nu}\bf{\nu'}}
(u_{\bf{\nu}}v_{\bf{\nu'}}+v_{\bf{\nu}}u_{\bf{\nu'}})\right\}
c^{\dagger}_{\bf{\nu}}c_{\bf{\nu'}}\nonumber \\
& & 
+\left\{-\varepsilon_{\bf{\nu}\bf{\nu'}}
u_{\bf{\nu}}v_{\bf{\nu'}}
+\frac{1}{2}N_{0}V^{00}_{\bf{\nu}\bf{\nu'}}
(u_{\bf{\nu}}u_{\bf{\nu'}}+v_{\bf{\nu}}v_{\bf{\nu'}})\right\}
(c^{\dagger}_{\bf{\nu}}c^{\dagger}_{\bf{\nu'}}+c_{\bf{\nu}}c_{\bf{\nu'}})
\biggl],
\end{eqnarray}
where $\varepsilon_{\bf{\nu}\bf{\nu'}}$ is given as
\begin{eqnarray}
\varepsilon_{\bf{\nu}\bf{\nu'}} &=& 
T_{\bf{\nu\nu'}}-\mu\delta_{\bf{\nu\nu'}}+
2\left(N_{0}V^{00}_{\bf{\nu\nu'}}
+n'_{\bf{\lambda}}V^{\bf{\lambda\nu}}_{\bf{\lambda\nu'}}\right).
\end{eqnarray}
We can eliminate the terms proportional to 
$c^{\dagger}_{\bf{\nu}}c^{\dagger}_{\bf{\nu'}}+c_{\bf{\nu}}c_{\bf{\nu'}}$ by 
imposing the condition
\begin{eqnarray}
-\varepsilon_{\bf{\nu}\bf{\nu'}}
u_{\bf{\nu}}v_{\bf{\nu'}}
+\frac{1}{2}N_{0}V^{00}_{\bf{\nu\nu'}}
(u_{\bf{\nu}}u_{\bf{\nu'}}+v_{\bf{\nu}}v_{\bf{\nu'}}) &=& 0,
\end{eqnarray}
or
\begin{eqnarray}
\varepsilon_{\bf{\nu\nu'}}u_{\bf{\nu'}}
-N_{0}V^{00}_{\bf{\nu\nu'}}v_{\bf{\nu'}} &=& 
E_{\bf{\nu'}}u_{\bf{\nu'}}\delta_{\nu\nu'},\\
\varepsilon_{\bf{\nu\nu'}}v_{\bf{\nu'}}
-N_{0}V^{00}_{\bf{\nu\nu'}}u_{\bf{\nu'}} &=& 
-E_{\bf{\nu'}}v_{\bf{\nu'}}\delta_{\nu\nu'},
\end{eqnarray}
where the eigenvalue $E_{\bf{\nu}}$ is given as
\begin{eqnarray}
E_{\bf{\nu}} = \sqrt{\varepsilon_{\bf{\nu\nu}}^{2}-
\left(N_{0}V^{00}_{\bf{\nu\nu}}\right)^{2}}.
\end{eqnarray}
Then, we obtain the diagonal Hamiltonian
\begin{eqnarray}
\hat{K} &=& (T_{00}-\mu)N_{0}+\frac{1}{2}V^{00}_{00}N^{2}_{0} 
+\sum_{{\bf \nu}\neq 0}E_{\bf{\nu}}c^{\dagger}_{\bf{\nu}}c_{\bf{\nu}}.
\end{eqnarray}
Here, we note that the first two terms in the right hand side of Eq. (31) 
correspond to the GP equation, and the third term in the right hand side of 
Eq. (25) represents small correction terms, and therefore we can ignore the 
third term. From Eq. (31), we can also obtain the thermodynamic potential
\begin{eqnarray}
\Omega &=& -k_{B}T\ln {\rm Tr} e^{-\hat{K}/k_{B}T}\nonumber\\
&\simeq&
(T_{00}-\mu)N_{0}
+\frac{1}{2}V^{00}_{00}N^{2}_{0}
+k_{B}T\sum_{{\bf \nu}\neq 0}\ln(1-e^{-E_{\bf{\nu}}/k_{B}T}).
\end{eqnarray}
The chemical potential is given by the condition
\begin{eqnarray}
0 &=& \frac{\partial \Omega}{\partial N_{0}}, \nonumber \\
&=& T_{00}-\mu+V^{00}_{00}N_{0}
+2\sum_{{\bf \nu} \neq 0} n'_{\bf{\nu}} V^{00}_{\bf{\nu\nu}},
\end{eqnarray}
where we ignore the anomalous average.\\
We note that Eqs. (31) and (33) are similar to results of the HFB 
approach to the Popov approximation \cite{gri}. In the next section, 
we will see the ground state stability of the interacting Bose gas 
using the GP equation.

\section{Ground State Stability}
In this section, we study the ground state properties of the Bose gas 
interacting through the Van der Waals type $-C_{6}r^{-6}$ and the 
$C_{3}r^{-3}$ interactions. We note that we must introduce a cut off at 
$r=2\langle R \rangle$, since these interactions diverge at $r=0$. Here, we 
assume the hard core potential inside the atomic radius $\langle R \rangle$. 
This is chosen to be the exponential or $C_{12}r^{-12}$ potential \cite{mwjk}.
\\
Now, we study the ground state stability of Bose gas for the condensate state 
at $T\le T_{c}$. In this case, the HFB theory is reliable for the 
investigation of the condensate state. To investigate the ground state 
stability, we employ a variational method. It is a good approximate scheme to 
study the ground state stability. We assume the following Gaussian wave 
function for the ground state
\begin{eqnarray}
\Psi({\bf r}) &=& \sqrt{\frac{N}{\pi^{3/2}\sigma^3}}e^{-r^2/2\sigma^2},
\end{eqnarray}
where $\sigma$ represents the variational parameter. This choice is natural 
when we take the noninteracting limit. 
\\
First, we consider the Van der Waals interaction. Here, we make comments on the 
scattering length $a$ for this interaction. This can be analytically given by 
\cite{gf}
\begin{eqnarray}
a &=&
\frac{\Gamma(3/4)}{2\sqrt{2}\Gamma(5/4)}
\left(\frac{mC_{6}}{\hbar^2}\right)^{\frac{1}{4}}
 \left[1-\tan\left(\phi-\frac{\pi}{8}\right)\right],
\end{eqnarray}
where $\phi$ is the semiclassical phase calculated at zero energy from the 
classical turning point to infinity. This phase depends on the repulsive hard 
core potential. Substituting Eq. (34) into Eq. (31), we can obtain the ground 
state energy $E_{g_6}$
\begin{eqnarray}
E_{g_6} &=& \frac{3}{4}N\hbar\omega(\sigma^2+\sigma^{-2})-
\frac{8N^2 C_{6}}{\pi a_{ho}^6 \sigma^{6}}I_{6},\nonumber\\
&=& \frac{3}{2}N\hbar\omega\left\{\frac{1}{2}(\sigma^2+\sigma^{-2})
-g_6\sigma^{-6}\right\},
\end{eqnarray}
where we rewrite $\sigma$ as $\sigma=\sigma/a_{ho}$ in units of the harmonic 
oscillator length $a_{ho}=\sqrt{\hbar/m\omega}$ and $I_{6}$ denotes the 
dimensionless integral
\begin{eqnarray}
I_{6} &=& 2\int^{\infty}_{0}ds\ s^2e^{-s^2}\int^{\infty}_{s+2\langle \tilde{R} 
\rangle}dt\ 
t^2e^{-t^2}\frac{1}{t^6}\frac{1+s^2/t^2}{(1-s^2/t^2)^4}.
\end{eqnarray}
The dimensionless coupling constant $g_{6}$ is defined by
\begin{eqnarray}
g_{6} &=& \frac{16NC_{6}I_{6}}{3\pi \hbar\omega a_{ho}^6}.
\end{eqnarray}
\begin{center}
\begin{tabular}{ccccc}
\multicolumn{5}{l}{Table 1. We plot the integral values }  \\
\multicolumn{5}{l}{of $I_{3}$ and $I_{6}$ for several cases 
of 2$\langle \tilde{R} \rangle$.} \\
 \hline\hline
$2\langle {\tilde R} \rangle $ & & $I_{3}$ & & $I_{6}$  \\ \hline
$1.0\times10^{-4}$  & $ $ & 1.295 & $ $ & 1.306$\times 10^{10}$  \\
$2.0\times10^{-4}$  & $ $ & 1.187 & $ $ & 1.632$\times 10^{9} $  \\
$3.0\times10^{-4}$  & $ $ & 1.123 & $ $ & 4.835$\times 10^{8} $  \\
$4.0\times10^{-4}$  & $ $ & 1.078 & $ $ & 2.040$\times 10^{8} $  \\
$5.0\times10^{-4}$  & $ $ & 1.043 & $ $ & 1.044$\times 10^{8} $  \\
$6.0\times10^{-4}$  & $ $ & 1.015 & $ $ & 6.044$\times 10^{7} $  \\
$7.0\times10^{-4}$  & $ $ & 0.991 & $ $ & 3.806$\times 10^{7} $  \\
$8.0\times10^{-4}$  & $ $ & 0.970 & $ $ & 2.550$\times 10^{7} $  \\
$9.0\times10^{-4}$  & $ $ & 0.951 & $ $ & 1.791$\times 10^{7} $  \\
$1.0\times10^{-3}$  & $ $ & 0.935 & $ $ & 1.306$\times 10^{7} $  \\
\hline\hline \\
\end{tabular}
\end{center}
In Table 1, we plot the values of $I_{6}$ as a function of the dimensionless 
parameter 2$\langle \tilde{R} \rangle = 2\langle R \rangle /a_{ho}$. 
We note that 2$\langle R \rangle$ and $a_{ho}$ are of the order of $\AA$ and 
$\mu$m, respectively. Therefore, 2$\langle \tilde{R} \rangle$ is of the order 
of $10^{-4}$. 
\\
Now, we look for the critical coupling constant ${\tilde g}_{6}$. This is given 
by the inflexion point of $E_{g_{6}}$
\begin{eqnarray}
\frac{d E_{g_{6}}}{d {\sigma}} = \frac{d^2 E_{g_{6}}}{d {\sigma}^2} = 0,
\end{eqnarray}
at ${\sigma}={\sigma}_{c}$ and $g_{6}={\tilde g_{6}}$. From Eq. (36), we obtain
\begin{eqnarray}
{\tilde g}_{6} &=& \frac{1}{24} \sim 0.0417.
\end{eqnarray}
Next, we consider the $C_3r^{-3}$ interaction \cite{grp}. We can write down 
the ground state energy $E_{g_3}$ 
\begin{eqnarray}
E_{g_3} &=& \frac{3}{4}N\hbar\omega(\sigma^2+\sigma^{-2})+
\frac{8N^2C_3}{\pi a_{ho}^3 \sigma^{3}}I_{3},\nonumber\\
&=& \frac{3}{2}N\hbar\omega\left\{\frac{1}{2}(\sigma^2+\sigma^{-2})
+g_3\sigma^{-3}\right\},
\end{eqnarray}
where $I_{3}$ is also a dimensionless integral defined as
\begin{eqnarray}
I_{3} &=& 2\int^{\infty}_{0}ds\ s^2e^{-s^2}\int^{\infty}_{s+2\langle 
\tilde{R} \rangle}dt\ t^2e^{-t^2}\frac{1}{t^3(1-s^2/t^2)}
\end{eqnarray}
In Table 1, we also plot the values of $I_{3}$ as a function of 
$2\langle\tilde{R}\rangle$. The dimensionless coupling constant $g_{3}$ is 
defined by
\begin{eqnarray}
g_{3} &=& \frac{16NC_{3}I_{3}}{3\pi\hbar\omega a_{ho}^3}.
\end{eqnarray}
In this case, the ground state energy has the same property as the contact 
interaction \cite{bay}. We can obtain the critical coupling constant 
${\tilde g}_{3}\sim -0.178$.
\\
Finally, we consider the $C_{3}r^{-3}-C_{6}r^{-6}$ interaction. Since the Van 
der Waals interaction is weak, we always assume $|g_{3}|>g_{6}\ge 0$. The ground 
state energy $E_{g}$ is given by
\begin{eqnarray}
E_{g} &=& \frac{3}{2}N\hbar\omega\left\{\frac{1}{2}(\sigma^2+\sigma^{-2})
+g_3\sigma^{-3}-g_6\sigma^{-6}\right\}.
\end{eqnarray}
\begin{figure}[htbp]
\begin{center}
\scalebox{0.5}[0.5]{\includegraphics{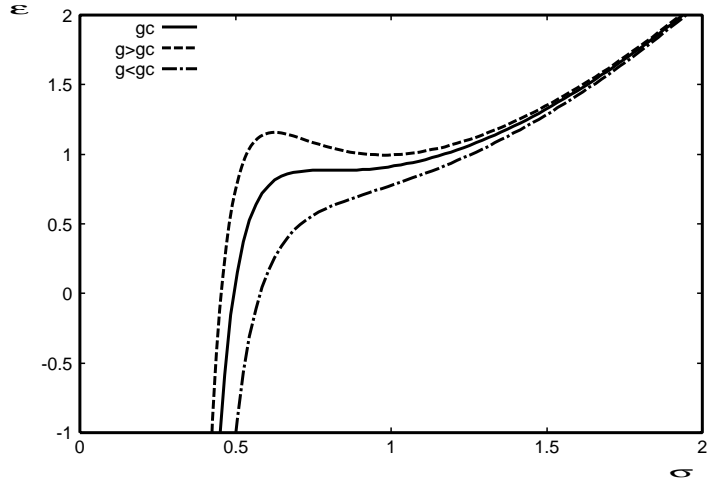}}
\caption{We show the ground state energy Eq. (44) in units of $3/2N\hbar\omega$ 
as a function of $\sigma$ for several values of the parameter $g=g_{3}/g_{6}$. 
The solid line is drawn with $g=g_{c}$. The dashed and dot-dashed lines are 
$g>g_{c}$ and $g<g_{c}$, respectively.}
\label{fig:fig1}
\end{center}
\end{figure}
In Fig. 1, we show the ground state energy in units of $3/2N\hbar\omega$ as a 
function of $\sigma$ for several values of the parameter $g=g_{3}/g_{6}$ which 
is given as
\begin{eqnarray}
g &=& \frac{C_{3}I_{3}}{C_{6}I_{6}}a_{ho}^3.
\end{eqnarray}
As can be seen, for $g>g_{c}$, the ground state energy always has a minimum. 
Therefore, the ground state is stable. On the other hand, for $g<g_{c}$, a 
minimum disappears. In this case, the ground state is unstable, therefore, BEC 
collapses. We can also calculate the critical value 
$g_{c}={\tilde g_{3}}/{\tilde g_{6}}$. It is given by
\begin{eqnarray}
9\tilde{g}_{3} &=& 8\sigma_{c}^5-4\sigma_{c},\\
18\tilde{g}_{6} &=& 5\sigma_{c}^8-\sigma_{c}^4.
\end{eqnarray}
Coupling constants $\tilde{g}_{3}$ and $\tilde{g}_{6}$ are related to each 
other through Eqs. (46) and (47). We note that the parameter $\sigma_{c}$ must 
be $\sigma_{c}\ge 0.66874$, since we assume $g_{6}\ge 0$. Eliminating $\sigma_{c}$ 
from Eqs. (46) and (47), we obtain
\begin{eqnarray}
\tilde{g}_{3} &=& 
\frac{8}{9}\left(\frac{1+\sqrt{1+360\tilde{g}_{6}}}{10}\right)^{5/4}-
\frac{4}{9}\left(\frac{1+\sqrt{1+360\tilde{g}_{6}}}{10}\right)^{1/4}.
\end{eqnarray}
\begin{figure}[htbp]
\begin{center}
\scalebox{0.5}[0.5]{\includegraphics{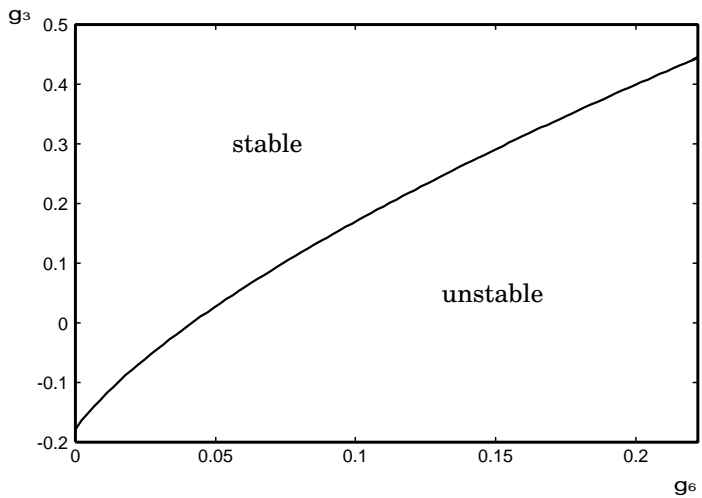}}
\caption{We show the phase diagram of coupling constants $g_{3}$ and $g_{6}$. 
The curve corresponds to the critical curve given by Eq. (48). The parameter 
runs $0.66874\le\sigma_{c}\le 1$ in this diagram. Above the critical curve, 
the ground state energy is stable, and below the critical curve, it is unstable.}
\label{fig:fig1}
\end{center}
\end{figure}
In Fig. 2, we show the phase diagram of coupling constants $g_{3}$ and 
$g_{6}$. The curve corresponds to the critical curve 
given by Eq. (48). Above the critical curve, the ground state energy is stable, 
and below the critical curve, it is unstable. Therefore, it is seen that the 
BEC collapses below the critical curve in the phase diagram.
\\
Now, we estimate the coupling constant $g$ for alkali atoms. Coefficients 
$C_{3}$ and $C_{6}$ are numerically given by Ref. \cite{md} for alkali atoms. 
\begin{center}
\begin{tabular}{ccccccccccc}
\multicolumn{11}{l}{Table 2. We plot the ratio $|C_{3}|/C_{6}\times10^3$
between $S$ and }  \\
\multicolumn{11}{l}{$P\sigma$ and$P\pi$ state for alkali 
atoms in units of the atomic} \\
\multicolumn{11}{l}{units (a.u.).} \\ \hline\hline
   $ $         & & Li    & & Na    & & K     & & Rb    & & Cs    \\ \hline
$S$-$P\sigma$  & & 5.329 & & 2.995 & & 1.845 & & 1.527 & & 1.205 \\
$S$-$P\pi$     & & 3.298 & & 2.325 & & 1.377 & & 1.144 & & 0.885 \\
\hline\hline \\
\end{tabular}
\end{center}
In Table 2, we plot the ratio $|C_{3}|/C_{6}$ for alkali atoms in units 
of the atomic units (a.u.). As can be seen from Table 2, $|C_{3}|/C_{6}$ is of 
the order of 10$^{-3}$ a.u.. The ratio $I_{3}/I_{6}$ 
and the harmonic oscillator length $a_{ho}$ are of the order of 
$10^{-7} \sim 10^{-8}$ and $10^{4}$ a.u., respectively. Thus, the coupling 
constant $g$ is of the order of $10 \sim 10^{2}$.  
\begin{figure}[htbp]
\begin{center}
\scalebox{0.5}[0.5]{\includegraphics{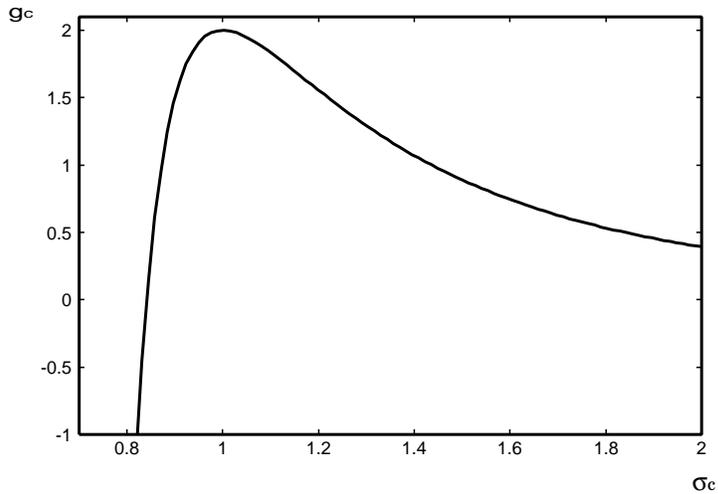}}
\caption{We show the critical coupling constant $g_{c}$ as a function of 
$\sigma_{c}$. Above the curve, the ground state energy is stable, and below the 
curve, it is unstable.}
\label{fig:fig1}
\end{center}
\end{figure}
In Fig. 3, we show the critical coupling constant $g_{c}$ as a function of 
$\sigma_{c}$. Above the curve, the ground state is stable, and below the curve, 
it is unstable. As can be seen from Fig. 3, for $g>0$, the ground state of 
alkali atoms are always stable in our calculation. On the other hand, for 
$g<0$, we must carefully consider the ground state stability, since it depends 
on $\sigma_{c}$. We can show that the ground state is stable when 
$\sigma_{c}\le 0.756$ for $g=-10$, and when $\sigma_{c}\le 0.690$ for 
$g=-10^2$. For real atoms, the value of $\sigma_{c}$ is expected to be the same 
as the $\delta$-function type interaction $\sigma_{c}=0.669$ \cite{bay}. 
Therefore, the ground state is also stable for $g<0$.
\\

\section{Conclusions}
We have studied the ground state stability of interacting Bose-Einstein gas 
with an arbitrary power low interaction.  In particular, we have considered 
the Van der Waals $-C_{6}r^{-6}$ plus $C_{3}r^{-3}$ interactions. Using the 
Gaussian variational function, we have obtained the ground state energy. It is 
shown that the ground state stability depends on the ratio of the coupling 
constants $g_{3}$ and $g_{6}$, then the critical coupling constant is obtained 
by the inflexion point of the ground state energy. For $g>g_{c}$, we can always 
produce stable BEC. On the other hand, for $g<g_{c}$, BEC collapses. Here, the 
ground state stability mainly depends on the $C_{3}r^{-3}$ term, since atoms 
become sensitive to this interaction when the density of atoms becomes high at 
low temperature. Therefore, we can understand that the role of this interaction 
is important for the stability of the ground state at low temperature. 
\\
Next, we have obtained the phase diagram of coupling constants $g_{3}$ and 
$g_{6}$. From the phase diagram, we have shown that there exists a stable 
region for Bose-Einstein gas. It is useful because we can classify atoms which 
become BEC.
\\
Finally, we have argued the validity of these results by using the numerical 
values of $C_{3}$ and $C_{6}$ for alkali atoms. We have shown that the ground 
state of alkali atoms are always stable.
\\
Our results show that we can reliably understand Bose-Einstein gas interacting 
through the realistic interaction beyond the GP theory.
\\

\section*{Acknowledgements}

I thank T. Fujita for careful reading.

\end{document}